\documentclass{pasj00}
\draft

\begin{document}
\SetRunningHead{S. Kato, A. Okazaki, and F. Oktariani}
               {Excitation of Disk Oscillations in Deformed Disks}
\Received{2010/0/00}
\Accepted{2010/12/28}

\title{Resonant Excitation of Disk Oscillations in Deformed Disks IV:  
        A New Formulation Studying Stability}

\author{Shoji \textsc{Kato}}%
\affil{2-2-2 Shikanodai-Nishi, Ikoma-shi, Nara 630-0114}
\email{kato.shoji@gmail.com; kato@kusastro.kyoto-u.ac.jp}

\author{Atsuo T. \textsc{Okazaki}}
\affil{Faculty of Engineering, Hokkai-Gakuen University, Toyohira-ku,
Sapporo 062-8605}
\author{Finny {\sc Oktariani}}
\affil{Department of Cosmoscience, Graduate School of Science, 
Hokkaido University,}
\affil{ Kita-ku, Sapporo 060-0810}

%


\maketitle

\begin{abstract}
The possibility has been suggested that high-frequency quasi-periodic oscillations observed in
low-mass X-ray binaries are resonantly excited disk oscillations in 
deformed (warped or eccentric) relativistic disks (Kato 2004).
In this paper we examine this wave excitation process from a viewpoint somewhat different
from that of previous studies.
We study how amplitudes of a set of normal mode oscillations change secularly with time by their
mutual couplings through disk deformation.
As a first step, we consider the case where the number of oscillation
modes contributing to the resonance coupling is two.
The results show that two prograde oscillations interacting through disk deformation can grow if 
their wave energies have opposite signs.
\end{abstract}


\section{Introduction}

The origin of high frequency quasi-periodic oscillations (HF QPOs) observed in low
mass X-ray binaries (LMXBs) is one of challenging subjects to be examined, since
its examination will clarify the structure of the innermost part of relativistic
accretion disks and the spin of central sources.
One promising possibility is that the QPOs are disk oscillations excited
in the innermost region of relativistic disks.
Particularly, the idea that HF QPOs are disk oscillations resonantly excited in deformed
(warped or eccentric) disks has been suggested by Kato (2004, 2008a,b) by analysical considerations, and
studied by Ferreria and Ogilvie (2008) and Oktariani et al. (2010) by numerical calculations.
In this model, a disk oscillation (hereafter, original oscillation) 
interacts non-linearly with the disk deformation  
to produce an oscillation (hereafter, intermediate oscillation).
The intermediate oscillation is a forced oscillation due to  
the coupling between the original oscillation and the deformation.
The intermediate oscillation then resonantly responds to the forcing
term at a certain radius (Lindblad resonance).
After having the resonance, the intermediate oscillation interacts non-linearly again with
the disk deformation to feed back to the original oscillation.
Through this feedback process the original and intermediate oscillations are excited, 
if cetain conditions are satisfied.

Some important consequences have been obtained so far in relation to the origin of the instability:
i) The wave energies of the original and intermediate oscillations must have opposite signs
(Kato 2004, 2008a).
That is, the instability is a result of wave energy exchange between two oscillations
with opposite signs of energy through a disk deformation.
ii) Since the intermediate oscillation resonantly responds to forcing terms resulting from the 
coupling between the disk deformation and the original oscillation,
its amplitude is not necessarily small compared with that of the original oscillation
(see figure 3 of Oktariani et al. 2010).
This means that the terminology of "original" and "intermediate" oscillations has
no particular meaning.
The two oscillations are equal partners, i.e.,
the role of the original oscillation mentioned in the previous 
paragraph can be also performed by the intermediate oscillation.
Hence, the interaction between two oscillations can be schematically sketched as figure 1.
This has been correctly acknowleged by Ferreira and Ogilvie (2008) (see figure 3 of their
paper).

Based on the above considerations, we develop here a perspective 
analytical method to study the instability.
In the previous analytical studies, we considered only the cases where the disk deformation 
is time-independent (i.e., its frequency of the disk deformation, $\omega_{\rm D}$, is zero) and frequencies of 
the two oscillations, $\omega_1$ and $\omega_2$, coupling through disk deformation are the same, i.e.,
$\omega_1=\omega_2$.
In the present analyses, the resonant condition is extended to $\omega_1=\omega_2\pm
\omega_{\rm D}$ with non-zero $\omega_{\rm D}$ and effects of weak deviation from the resonant condition, 
$\omega_1=\omega_2\pm\omega_{\rm D}$, on growth rate of oscillations are also examined. 

It is noted that the number of oscillations contributing to this resonant process is not always 
limited to two.
In order to understand the essence of this instability, however, we assume in this paper that 
only two modes of oscillations contribute to this resonsnt process.
More general cases will be a subject in the future.

\begin{figure}
\begin{center}
    \FigureFile(80mm,80mm){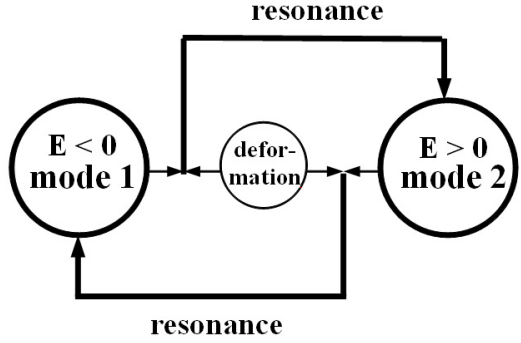}
\end{center}
\caption{Schematic diagram showing two resonant interactions between two disk oscillations
(mode 1 and mode 2) through couplings with disk deformation.
Two oscillations must have opposite signs in their wave energy.
In this figure, mode 1 is taken to have a negative energy} 
\end{figure}

\section{Basic Hydrodynamical Equations and Some Other Relations}

We summarize here basic equations and relations [see Kato (2008a, b) for details]
to be used to discuss wave couplings through a disk deformation.

\subsection{Nonlinear Hydrodynamical Equations}

When we try to quantitatively apply the present excitation process to high-frequency QPOs 
observed in GBHCs (galactic black hole candidates) and LMXBs,
effects of general relativity should be taken into account.
The general relativity is, however, not essential to understanding the essence of the 
instability mechanism.
Hence, in this paper, for simplicity, formulation is done in  
the framework of a pseudo-Newtonian potential, using the 
gravitational potential introduced by Paczy{\' n}ski and Wiita (1980).
We adopt a Lagrangian formulation by Lynden-Bell and Ostriker (1967).

The unperturbed disk is in a steady equilibrium state.
Over the equilibrium state, weakly nonlinear perturbations are superposed.
By using a displacement vector, $\mbox{\boldmath $\xi$}$, 
the weakly nonlinear hydrodynamical equation describing adiabatic, 
nonself-gravitating perturbations is written as, after lengthly manipulation 
(Lynden-Bell and Ostriker 1967),
\begin{equation}
    \rho_0{\partial^2\mbox{\boldmath $\xi$}\over\partial t^2}
         +2\rho_0(\mbox{\boldmath $u$}_0\cdot\nabla)
                 {\partial\mbox{\boldmath $\xi$}\over\partial t}
      +\mbox{\boldmath $L$}(\mbox{\boldmath $\xi$})
      =\rho_0\mbox{\boldmath $C$}(\mbox{\boldmath $\xi$},
                                                       \mbox{\boldmath $\xi$}),
\label{2.3}
\end{equation}
where $\mbox{\boldmath $L$}(\mbox{\boldmath $\xi$})$ is a linear
Hermitian operator with respect to $\mbox{\boldmath $\xi$}$ and is
\begin{eqnarray}
   \mbox{\boldmath $L$}(\mbox{\boldmath $\xi$})
     =\rho_0(\mbox{\boldmath $u$}_0\cdot\nabla)
                         (\mbox{\boldmath $u$}_0\cdot\nabla)\mbox{\boldmath $\xi$}
       +\rho_0(\mbox{\boldmath $\xi$}\cdot\nabla)(\nabla\psi_0)
       +\nabla\biggr[(1-\Gamma_1)p_0{\rm div}\mbox{\boldmath $\xi$}\biggr] 
                          \nonumber \\
       -p_0\nabla({\rm div}\mbox{\boldmath $\xi$})
       -\nabla[(\mbox{\boldmath $\xi$}\cdot\nabla)p_0]
       +(\mbox{\boldmath $\xi$}\cdot\nabla)(\nabla p_0),
\label{2.4}
\end{eqnarray}
and $\rho_0(\mbox{\boldmath $r$})$ and $p_0(\mbox{\boldmath $r$})$ are respectively 
the density and pressure in the unperturbed state, and $\Gamma_1$ is
the barotropic index specifying the linear part of the relation between 
Lagrangian variations $\delta p$ and $\delta \rho$, i.e.,
$(\delta p/p_0)_{\rm linear}=\Gamma_1(\delta\rho/\rho_0)_{\rm linear}$.
Since the self-gravity of the disk gas has been neglected,
the gravitational potential, $\psi_0(\mbox{\boldmath $r$})$, is a given function and
has no Eulerian perturbation. 
In the above hydrodynamical equations (\ref{2.3}) and (\ref{2.4}), there is no restriction
on the form of unperturbed flow $\mbox{\boldmath $u$}_0$.
However, in the followings we assume that the unperturbed flow is a cylindrical rotation
alone, i.e., $\mbox{\boldmath $u$}_0 = (0,r\Omega,0)$, in the cylindrical
cordinates ($r$, $\varphi$, $z$), where the origin is at the disk center and the $z$-axis 
is in the direction perpendicular to the unperturbed disk plane with 
$\Omega(r)$ being the angular velocity of disk rotation.

The right-hand side of wave equation (\ref{2.3}) represents the weakly nonlinear terms.
No detailed expression for $\mbox{\boldmath $C$}$ is given here
(for detailed expressions, see Kato 2004, 2008a), but an important characteristics of 
$\mbox{\boldmath $C$}$ is that we have commutative relations (Kato 2008a)
for an arbitrary set of $\mbox{\boldmath $\eta$}_1$, 
$\mbox{\boldmath $\eta$}_2$, and $\mbox{\boldmath $\eta$}_3$, e.g.,
\begin{equation}
    \int\rho_0{\mbox{\boldmath $\eta$}}_1
                             \mbox{\boldmath $C$}
                          ({\mbox{\boldmath $\eta$}}_2, 
                           {\mbox{\boldmath $\eta$}}_3)dV
        =\int\rho_0{\mbox{\boldmath $\eta$}}_1
                             \mbox{\boldmath $C$}
                          ({\mbox{\boldmath $\eta$}}_3, 
                           {\mbox{\boldmath $\eta$}}_2)dV
        =\int\rho_0{\mbox{\boldmath $\eta$}}_3
                              \mbox{\boldmath $C$}
                           ({\mbox{\boldmath $\eta$}}_1, 
                            {\mbox{\boldmath $\eta$}}_2)dV.
\label{commutative}
\end{equation}
As shown later, the presence of these commutative relations leads to a simple expression 
of instability criterion.
We suppose that the presence of these commutative relations is a general property of conservative systems
beyond the assumption of weak nonlinearity.

\subsection{Orthogonality of Normal Modes}

In preparation for subsequent studies, some orthogonality relations are summarized here.
Eigen-functions describing linear oscillations in non-deformed disks are denoted by 
$\mbox{\boldmath $\xi$}_\alpha(\mbox{\boldmath $r$},t)$.
Here, the subscript $\alpha$ is used to distinct all eigen-functions.
Time-dependent part of $\mbox{\boldmath $\xi$}_\alpha(\mbox{\boldmath $r$},t)$
is expressed as exp$(i\omega_\alpha t)$, where $\omega_\alpha$ is real.
Then, $\mbox{\boldmath $\xi$}_\alpha(\mbox{\boldmath $r$},t)$ satisfies
\begin{equation}
    -\omega_\alpha^2\rho_0\mbox{\boldmath $\xi$}_\alpha
         +2i\omega_\alpha\rho_0(\mbox{\boldmath $u$}_0\cdot\nabla)
                 \mbox{\boldmath $\xi$}_\alpha
      +\mbox{\boldmath $L$}(\mbox{\boldmath $\xi$}_\alpha)=0.
\label{2.5}
\end{equation}
Now, this equation is multiplied by $\mbox{\boldmath $\xi$}_\beta^*(\mbox{\boldmath $r$},t)$
and integrated over the whole volume, where the superscript * denotes the complex conjugate and 
$\beta\not=\alpha$.
The volume integration of $\rho_0\mbox{\boldmath $\xi$}_\beta^*(\mbox{\boldmath $r$},t)
\mbox{\boldmath $\xi$}_\alpha(\mbox{\boldmath $r$},t)$ over the whole volume is written 
hereafter as $\langle\rho_0\mbox{\boldmath $\xi$}_\beta^*\mbox{\boldmath $\xi$}_\alpha\rangle$.
Then, we have 
\begin{equation}
    -\omega_\alpha^{2}\langle\rho_0\mbox{\boldmath $\xi$}_\beta^*\mbox{\boldmath $\xi$}_\alpha\rangle
    +2i\omega_\alpha\langle \rho_0\mbox{\boldmath $\xi$}_\beta^*(\mbox{\boldmath $u$}_0\cdot\nabla)
       \mbox{\boldmath $\xi$}_\alpha\rangle
    +\langle\mbox{\boldmath $\xi$}_\beta^*\cdot\mbox{\boldmath $L$}(\mbox{\boldmath $\xi$}_\alpha)\rangle
    =0.  
\label{2.6}
\end{equation}
Similarly, integrating the linear wave equation of $\mbox{\boldmath $\xi$}_\beta^*$
over the whole volume after the equation being multiplied by
$\mbox{\boldmath $\xi$}_\alpha$, we have
\begin{equation}
    -\omega_\beta^{2}\langle\rho_0\mbox{\boldmath $\xi$}_\alpha\mbox{\boldmath $\xi$}_\beta^*\rangle
    -2i\omega_\beta\langle \rho_0\mbox{\boldmath $\xi$}_\alpha(\mbox{\boldmath $u$}_0\cdot\nabla)
       \mbox{\boldmath $\xi$}_\beta^*\rangle
    +\langle\mbox{\boldmath $\xi$}_\alpha\cdot\mbox{\boldmath $L$}(\mbox{\boldmath $\xi$}_\beta^*)\rangle
    =0.  
\label{2.7}
\end{equation}
Since the operator  $\mbox{\boldmath $L$}$ is a Hermitian (Lynden-Bel and Ostriker 1967),
we have the relation of
\begin{equation}
   \langle\mbox{\boldmath $\xi$}_\alpha\cdot\mbox{\boldmath $L$}(\mbox{\boldmath $\xi$}_\beta^*)\rangle
   =\langle[\mbox{\boldmath $L$}(\mbox{\boldmath $\xi$}_\alpha^*)]^*\cdot
      \mbox{\boldmath $\xi$}_\beta^*\rangle
   =\langle\mbox{\boldmath $L$}(\mbox{\boldmath $\xi$}_\alpha)\cdot
                         \mbox{\boldmath $\xi$}_\beta^*\rangle.
\label{2.8}
\end{equation}
Hence, the difference of the above two equations [eqs. (\ref{2.6}) and (\ref{2.7})] gives, 
when $\omega_\beta\not=\omega_\alpha$,
\begin{equation}
    (\omega_\alpha+\omega_\beta)\langle\rho_0\mbox{\boldmath $\xi$}_\alpha
       \mbox{\boldmath $\xi$}_\beta^*\rangle
        =2i\langle\rho_0\mbox{\boldmath $\xi$}_\beta^*(\mbox{\boldmath $u$}_0\cdot\nabla)
        \mbox{\boldmath $\xi$}_\alpha\rangle
        =-2i\langle\rho_0\mbox{\boldmath $\xi$}_\alpha(\mbox{\boldmath $u$}_0\cdot\nabla)
        \mbox{\boldmath $\xi$}_\beta^*\rangle.
\label{2.9}
\end{equation}
In deriving the last equality, we have used an integration by part, assuming that $\rho_0$ 
vanishes on the disk surface.

Different from the case of non-rotating stars, the eigenfunctions of normal modes of disk oscillations are
not orthogonal in the sense of $\langle\rho_0\mbox{\boldmath $\xi$}_\alpha
\mbox{\boldmath $\xi$}_\beta^*\rangle = 0$.
In spite of this, however, eigenfunctions of disk oscillations are orthogonal in many situations.
They are classified by azimuthal wavenumber, $m$, 
node number in the vertical direction, $n$, and that in the radial direction, $\ell$, 
in addition to the distinction of p- and g-modes [see Kato (2001) or Kato et al. (2008) for
classification of disk oscillations]. 
Eigenfunctions with different azimuthal wavenumber are obviously orthogonal,
i.e., $\langle\rho_0\mbox{\boldmath $\xi$}_\alpha
\mbox{\boldmath $\xi$}_\beta^*\rangle= 0$ when $m_\alpha\not= m_\beta$.
Even if the azimuthal wavenumbers are the same, $\langle\rho_0\mbox{\boldmath $\xi$}_\alpha
\mbox{\boldmath $\xi$}_\beta^*\rangle =0$ when $n_\alpha\not=n_\beta$, 
if the disk is geometrically thin and isothermal in the vertical direction.
This comes from the fact that in such disks the $z$-dependence of eigen-functions with $n$ node(s) in the
$z$-direction ($n$ is zero or a positive integer) is described by the Hermite
polynomials ${\cal H}_n$ as 
\begin{equation}
     \xi_r, \ \xi_\varphi\propto {\cal H}_n(z/H), \quad \xi_z\propto {\cal H}_{n-1}(z/H),
\label{2.10}
\end{equation}
(Okazaki et al. 1987), where the subscripts $r$, $\varphi$, and $z$ represent the cylindrical coordinates
($r$, $\varphi$, $z$) whose origin is at the disk center and 
the $z$-axis is perpendicular to the disk plane.
Here, ${\cal H}_n$ is the Hermite polynomial of argument $z/H$, 
$H$ being the half-thickness of the disk.
Thus, the eigen-functions classified by $m$ and $n$ are orthogonal.\footnote{
In vertically polytropic disks, orthogonality holds by using the Gegenbauer polynomials
(Silbergleit et al. 2001).
}
In summary, we have
\begin{equation}
    \langle\rho_0\mbox{\boldmath $\xi$}_\alpha\mbox{\boldmath $\xi$}_\beta^*\rangle
     =\langle\rho_0\mbox{\boldmath $\xi$}_\alpha\mbox{\boldmath $\xi$}_\beta^*\rangle
        \delta_{m_\alpha,m_\beta}\delta_{n_\alpha,n_\beta},
\label{2.11}
\end{equation}
where $\delta_{a,b}$ is the Kronecker delta, i.e., it is unity when $a=b$, while zero when $a\not= b$.

Orthogonality of $\langle\rho_0\mbox{\boldmath $\xi$}_\alpha
\mbox{\boldmath $\xi$}_\beta^*\rangle$ does not hold in the case where $m_\alpha=m_\beta$ and 
$n_\alpha=n_\beta$.
Even in these cases, however, $\langle\rho_0\mbox{\boldmath $\xi$}_\alpha
\mbox{\boldmath $\xi$}_\beta^*\rangle$ will be close to zero for $\ell_\alpha\not=\ell_\beta$, 
if our interest is on short wavelength oscillations in the radial 
direction, since the radial dependence of eigenfunctions is close to sinusoidal in such cases.

\section{Couplings of Two Oscillations through Disk Deformation}

Let us consider the case where two oscillation modes, 
$\mbox{\boldmath $\xi$}_1(\mbox{\boldmath $r$},t)$
and $\mbox{\boldmath $\xi$}_2(\mbox{\boldmath $r$},t)$, resonantly couple through disk deformation
$\mbox{\boldmath $\xi$}_{\rm D}(\mbox{\boldmath $r$},t)$.
Through the coupling term $\mbox{\boldmath $C$}(\mbox{\boldmath $\xi$},\mbox{\boldmath $\xi$})$ 
[see eq.(\ref{2.3})] many other modes than $\mbox{\boldmath $\xi$}_1$ and $\mbox{\boldmath $\xi$}_2$ 
appear, and their amplitudes as well as those of $\mbox{\boldmath $\xi$}_1$ and $\mbox{\boldmath $\xi$}_2$
become time dependent.
Now, we assume that normal modes of oscillations form a complete set, and
expand the resulting oscillations, $\mbox{\boldmath $\xi$}(\mbox{\boldmath $r$},t)$, 
including the disk deformation, $\mbox{\boldmath $\xi$}_{\rm D}(\mbox{\boldmath $r$},t)$,
in the form
\begin{equation}
   \mbox{\boldmath $\xi$}(\mbox{\boldmath $r$},t)=
    A_1(t)\mbox{\boldmath $\xi$}_1(\mbox{\boldmath $r$},t)
    +A_2(t)\mbox{\boldmath $\xi$}_2(\mbox{\boldmath $r$},t)
    +A_{\rm D}\mbox{\boldmath $\xi$}_{\rm D}(\mbox{\boldmath $r$},t)
    +\sum_\alpha A_\alpha(t)\mbox{\boldmath $\xi$}_\alpha(\mbox{\boldmath $r$},t).
\label{3.1}
\end{equation}
Since our main concern is on the modes 1 and 2, 
$\mbox{\boldmath $\xi$}_1$ and $\mbox{\boldmath $\xi$}_2$ are distinguished from other eigen-functions
and the subscript $\alpha$ is hereafter used only to denote other eigenfunctions than 
$\mbox{\boldmath $\xi$}_1$ and $\mbox{\boldmath $\xi$}_2$.
Our purpose here is to derive equations describing a secular time evolution of $A_1$ and $A_2$.
The disk deformation $\mbox{\boldmath $\xi$}_{\rm D}(\mbox{\boldmath $r$},t)$ is assumed to have 
a much larger amplitude than other oscillations and its time variation during the coupling
processes is neglected, i.e., $A_{\rm D}=$ const.

We now express the eigen-frequencies associated with $\mbox{\boldmath $\xi$}_1$,
$\mbox{\boldmath $\xi$}_2$, $\mbox{\boldmath $\xi$}_{\rm D}$, and $\mbox{\boldmath $\xi$}_\alpha$
by $\omega_1$, $\omega_2$, $\omega_{\rm D}$, and $\omega_\alpha$, respectively, i.e., 
$\mbox{\boldmath $\xi$}_1(\mbox{\boldmath $r$},t)={\rm exp}(i\omega_1 t)
\mbox{\boldmath $\xi$}_1(\mbox{\boldmath $r$})$ etc.
Then, substitution of equation (\ref{3.1}) into equation (\ref{2.3}) leads to
\begin{eqnarray}
  &&  2\rho_0\frac{dA_1}{dt}[i\omega_1+(\mbox{\boldmath $u$}_0\cdot\nabla)]\mbox{\boldmath $\xi$}_1
   +2\rho_0\frac{dA_2}{dt}[i\omega_2+(\mbox{\boldmath $u$}_0\cdot\nabla)]\mbox{\boldmath $\xi$}_2 
   +\sum_\alpha 2\rho_0\frac{dA_\alpha}{dt}[i\omega_\alpha+(\mbox{\boldmath $u$}_0\cdot\nabla)]\mbox{\boldmath 
   $\xi$}_\alpha \nonumber \\
  && = \sum_{i=1,2}\frac{1}{2}A_i\biggr[A_{\rm D}\biggr(\rho_0\mbox{\boldmath $C$}(\mbox{\boldmath $\xi$}_i,\mbox{\boldmath $\xi$}_{\rm D})
   +\rho_0\mbox{\boldmath $C$}(\mbox{\boldmath $\xi$}_{\rm D},\mbox{\boldmath $\xi$}_i)\biggr)
   +A_{\rm D}^*\biggr(\rho_0\mbox{\boldmath $C$}(\mbox{\boldmath $\xi$}_i,\mbox{\boldmath $\xi$}_{\rm D}^*)
   +\rho_0\mbox{\boldmath $C$}(\mbox{\boldmath $\xi$}_{\rm D}^*,\mbox{\boldmath $\xi$}_i)\biggr)\biggr] 
    \nonumber \\
  && +\sum_\alpha\frac{1}{2}A_\alpha\biggr[A_{\rm D}\biggr(\rho_0\mbox{\boldmath $C$}(\mbox{\boldmath $\xi$}_\alpha,\mbox{\boldmath $\xi$}_{\rm D})
     +\rho_0\mbox{\boldmath $C$}(\mbox{\boldmath $\xi$}_{\rm D},\mbox{\boldmath $\xi$}_\alpha)\biggr)
   +A_{\rm D}^*\biggr(\rho_0\mbox{\boldmath $C$}(\mbox{\boldmath $\xi$}_\alpha,\mbox{\boldmath $\xi$}_{\rm D}^*)
     +\rho_0\mbox{\boldmath $C$}(\mbox{\boldmath $\xi$}_{\rm D}^*,\mbox{\boldmath $\xi$}_\alpha)\biggr)
   \biggr],
\label{3.2}
\end{eqnarray}
where terms of $d^2A_1/dt^2$, $d^2A_2/dt^2$, and $d^2A_\alpha/dt^2$ 
have been neglected, since we are interested in slow secular evolutions of $A'$s.
On the right-hand side of equation (\ref{3.2}), the coupling terms that 
are not related to the disk deformation are neglected.\footnote{
In writing down the right-hand side of equation (\ref{3.2}), we have used the following relation:
\begin{eqnarray}
     \Re(A)\Re(B)=\frac{1}{2}\Re[AB+AB^*],     \nonumber
\end{eqnarray}
where $A$ and $B$ are complex variables.}

Now, we define the wave energy $E_1$ of normal mode of oscillation $\mbox{\boldmath $\xi$}_1$ by 
\begin{equation}
   E_1=\frac{1}{2}\omega_1\biggr[\omega_1\langle\rho_0\mbox{\boldmath $\xi$}_1^*
            \mbox{\boldmath $\xi$}_1\rangle
       -i\langle\rho_0\mbox{\boldmath $\xi$}_1^*(\mbox{\boldmath $u$}_0\cdot\nabla)
            \mbox{\boldmath $\xi$}_1\rangle\biggr]
\label{3.4}
\end{equation}
(Kato 2001, 2008a).
To use later, the wave energy $E_2$ of the normal mode $\mbox{\boldmath $\xi$}_2$ is also
defined by
\begin{equation}
   E_2=\frac{1}{2}\omega_2\biggr[\omega_2\langle\rho_0\mbox{\boldmath $\xi$}_2^*
            \mbox{\boldmath $\xi$}_2\rangle
       -i\langle\rho_0\mbox{\boldmath $\xi$}_2^*(\mbox{\boldmath $u$}_0\cdot\nabla)
            \mbox{\boldmath $\xi$}_2\rangle\biggr].
\label{3.4'}
\end{equation}
Furthermore, we introduce the following quantities:
\begin{equation}
       W_{11}= \frac{1}{2}\biggr(\biggr\langle\rho_0\mbox{\boldmath $\xi$}_1^*\cdot
       \mbox{\boldmath $C$}(\mbox{\boldmath $\xi$}_1, \mbox{\boldmath $\xi$}_{\rm D})\biggr\rangle
       +\biggr\langle\rho_0\mbox{\boldmath $\xi$}_1^*\cdot\mbox{\boldmath $C$}(\mbox{\boldmath $\xi$}_{\rm D}, 
                 \mbox{\boldmath $\xi$}_1)\biggr\rangle\biggr),
\label{W11}
\end{equation}
\begin{equation}       
       W_{11*}= \frac{1}{2}\biggr(\biggr\langle\rho_0\mbox{\boldmath $\xi$}_1^*
       \cdot\mbox{\boldmath $C$}(\mbox{\boldmath $\xi$}_1, \mbox{\boldmath $\xi$}_{\rm D}^*)\biggr\rangle
       +\biggr\langle\rho_0\mbox{\boldmath $\xi$}_1^*
       \cdot\mbox{\boldmath $C$}(\mbox{\boldmath $\xi$}_{\rm D}^*, \mbox{\boldmath $\xi$}_1)\biggr\rangle\biggr),
\label{W11'}
\end{equation}
\begin{equation}
       W_{12}= \frac{1}{2}\biggr(\biggr\langle\rho_0\mbox{\boldmath $\xi$}_1^*\cdot
       \mbox{\boldmath $C$}(\mbox{\boldmath $\xi$}_2, \mbox{\boldmath $\xi$}_{\rm D})\biggr\rangle
       +\biggr\langle\rho_0\mbox{\boldmath $\xi$}_1^*\cdot\mbox{\boldmath $C$}(\mbox{\boldmath $\xi$}_{\rm D}, 
                 \mbox{\boldmath $\xi$}_2)\biggr\rangle\biggr),
\label{W12}
\end{equation}
\begin{equation}       
       W_{12*}= \frac{1}{2}\biggr(\biggr\langle\rho_0\mbox{\boldmath $\xi$}_1^*
       \cdot\mbox{\boldmath $C$}(\mbox{\boldmath $\xi$}_2, \mbox{\boldmath $\xi$}_{\rm D}^*)\biggr\rangle
       +\biggr\langle\rho_0\mbox{\boldmath $\xi$}_1^*
       \cdot\mbox{\boldmath $C$}(\mbox{\boldmath $\xi$}_{\rm D}^*, \mbox{\boldmath $\xi$}_2)\biggr\rangle\biggr),
\label{W12'}
\end{equation}
\begin{equation}
       W_{1\alpha}= \frac{1}{2}\biggr(\biggr\langle\rho_0\mbox{\boldmath $\xi$}_1^*\cdot
       \mbox{\boldmath $C$}(\mbox{\boldmath $\xi$}_\alpha, \mbox{\boldmath $\xi$}_{\rm D})\biggr\rangle
       +\biggr\langle\rho_0\mbox{\boldmath $\xi$}_1^*\cdot\mbox{\boldmath $C$}(\mbox{\boldmath $\xi$}_{\rm D}, 
        \mbox{\boldmath $\xi$}_\alpha)\biggr\rangle\biggr),
\label{W1alpha}
\end{equation}
\begin{equation}       
       W_{1\alpha *}= \frac{1}{2}\biggr(\biggr\langle\rho_0\mbox{\boldmath $\xi$}_1^*
       \cdot\mbox{\boldmath $C$}(\mbox{\boldmath $\xi$}_\alpha, \mbox{\boldmath $\xi$}_{\rm D}^*)\biggr\rangle
       +\biggr\langle\rho_0\mbox{\boldmath $\xi$}_1^*
       \cdot\mbox{\boldmath $C$}(\mbox{\boldmath $\xi$}_{\rm D}^*, \mbox{\boldmath $\xi$}_\alpha)\biggr\rangle\biggr).
\label{W1alpha'}
\end{equation}

To proceed further, the time and azimuthal dependences of normal mode oscillations are
written explicitly as
\begin{equation}
     \mbox{\boldmath $\xi$}_k(\mbox{\boldmath $r$}, t)={\rm exp}[i(\omega_k t-m_k\varphi)]
        \hat {\mbox{\boldmath $\xi$}}_k  \quad (k=1,\ 2,\ \alpha,\ {\rm D}).
\label{3.41}
\end{equation}
To avoid writing down similar relations repeatedly, we have introduced the subscript $k$, 
which represents all oscillation modes, i.e., $k$ denotes 1, 2, $\alpha$ and ${\rm D}$.
Here, we take all $m_k$ to be zero or positive intergers, while $\omega_k$ is not always positive.
If $\omega_k<0$, the oscillation is retrograde.

It is noted here that by using equation (\ref{3.41}), we can express the wave energy of the normal mode
oscillation in an instructive form.
Since the $r$- and $\varphi$- components of $\mbox{\boldmath $\xi$}_1$, say $\xi_{1r}$ and $\xi_{1\varphi}$,
are related in a geometrically thin disks by (e.g., Kato 2004)
\begin{equation}
            i(\omega_1-m\Omega_1)\xi_{1\varphi}+2\Omega\xi_{1r}\sim 0,
\label{varphi}
\end{equation}
we have
\begin{equation}
        E_1\sim \frac{\omega_1}{2}\biggr\langle (\omega_1-m_1\Omega)\rho_0(\xi_{1r}^*\xi_{1r}+\xi_{1z}^*\xi_{1z})\biggr\rangle.
\label{energy}
\end{equation}
This shows that the sign of wave energy is determined by the sign of $\omega_1-m_1 \Omega$ in the region
where the wave exists predominantly (e.g., Kato 2001).
For example, a prograde ($\omega_1>0$) wave inside the corotation resonance has a negativce energy, 
while a prograde wave outside
it has a positive energy.

In previous papers we mainly considered the case of $\omega_1=\omega_2$  with $\omega_{\rm D}=0$.
In this paper, we extend our analyses to more general cases of resonance: 
\begin{equation}
      \omega_1\sim \omega_2\pm \omega_{\rm D},
\label{resonance}
\end{equation}
where $\omega_{\rm D}$ is not necessary to be small.
We introduce $\Delta_+$ and $\Delta_-$ defined by
\begin{equation}
      \Delta_+=\omega_1-\omega_2-\omega_{\rm D} \quad {\rm and}\quad \Delta_-=\omega_1-\omega_2+\omega_{\rm D}.
\label{delta}
\end{equation}
In the resonance of $\omega_1\sim\omega_2+\omega_{\rm D}$, $\Delta_+$ is small, but $\Delta_-$ is not small
unless $\omega_{\rm D}$ is small.
In the resonance of $\omega_1\sim\omega_2-\omega_{\rm D}$, on the other hand, $\Delta_-$ is small,
but $\Delta_+$ is not always so unless $\omega_{\rm D}$ is small. 
It is noted that the resonant condition concerning azimuthal wavenumber is
\begin{equation}
        m_1=m_2\pm m_{\rm D},
\label{m-resonance}
\end{equation}
where $m_1$ and $m_2$ are zero or positive integers, while $m_{\rm D}$ is a positive integer, since
we focus our attention only on the case where the disk deformation is non-axisymmetric, i.e., 
$m_{\rm D}\not= 0$.

After these preparations, we integrate equation (\ref{3.2}) over the whole volume after multiplying  
$\mbox{\boldmath $\xi$}_1^*(\mbox{\boldmath $r$},t)$. 
Then, the term with $dA_1/dt$ becomes $i(4E_1/\omega_1)(dA_1/dt)$, while the term with $dA_2/dt$ vanishes
since $m_2\not= m_1$ by definition.
Concerning the term with $dA_\alpha/dt$, some more consideration is necessary.
If $\omega_\alpha\not= \omega_1$, by using equation (\ref{2.9}) we can reduce the integration to 
\begin{equation}
     i\sum_\alpha\frac{dA_\alpha}{dt}(\omega_\alpha-\omega_1)\langle\rho_0\mbox{\boldmath $\xi$}_1^*
            \mbox{\boldmath $\xi$}_\alpha\rangle. 
\label{*}
\end{equation}
In the case where $m_\alpha=m_1\pm m_{\rm D}$, 
$\langle\rho_0\mbox{\boldmath $\xi$}_1^*\mbox{\boldmath $\xi$}_\alpha\rangle$ vanishes since $m_{\rm D}\not= 0$.
On the other hand, when $m_\alpha\not= m_1\pm m_{\rm D}$, $A_\alpha$ does not appear in the coupling term.
That is, $A_1$ and $A_\alpha$ have no nonlinear couplings and thus we can take as $A_\alpha=0$ when we consider
time evolution of $A_1$.
In the case of $\omega_\alpha=\omega_1$, the last term on the left-hand side of equation (\ref{3.2}) does 
not lead to equation (\ref{*}).
Even in this case, by the same arguments as the above we can neglect the term.
Considering these situations we have
\begin{eqnarray}
   &&i\frac{dA_1}{dt}\frac{4E_1}{\omega_1}  \nonumber \\
   &&=A_1(A_{\rm D}W_{11}+A_{\rm D}^*W_{11*})
    +A_2(A_{\rm D}W_{12}+A_{\rm D}^*W_{12*})
    +\sum_\alpha A_\alpha(A_{\rm D}W_{1\alpha}+A_{\rm D}^*W_{1\alpha *}).
\label{3.5}
\end{eqnarray}

Among various coupling terms on the right-hand side of equation (\ref{3.5}),
the first two terms with $W_{11}$ and $W_{11*}$ can be neglected if we consider non-axisymmetric disk
deformations such as warp or eccentric deformation, since the terms inside $\langle \ \ \rangle$
in equations (\ref{W11}) and (\ref{W11'}) are proportional to  
${\rm exp}(-im_{\rm D}\varphi)$ and ${\rm exp}(im_{\rm D}\varphi)$, respectively,
and their angular averages vanish.
The last two terms with $W_{1\alpha}$ and $W_{1\alpha *}$ on the right-hand side of 
equation (\ref{3.5}) are also neglected hereafter by the following reasons.
The terms inside of $\langle \ \ \rangle$ of equations (\ref{W1alpha}) and (\ref{W1alpha'})
are proportional to ${\rm exp}[i(-\omega_1+\omega_\alpha+\omega_{\rm D})t]$
and ${\rm exp}[i(-\omega_1+\omega_\alpha-\omega_{\rm D})t]$, respectively.
In general, they rapidly vary with time, since no resonant condition is assumed among
$\omega_1$, $\omega_\alpha$, and $\omega_{\rm D}$.
Hence, if short timescale variations are averaged over,\footnote{
We are interested in solutions where all $A$'s vary slowly with time
} 
the averaged quantities are small and
can be neglected.\footnote{
In some cases, however, some of $\omega_\alpha$'s is/are close to $\omega_1\pm\omega_{\rm D}$.
For example, all trapped axi-symmetric g-mode oscillations have 
frequencies close to $\kappa_{\rm max}$
(the maximum of the epicyclic frequency), when their azimuthal wavenumber is zero.
Then, $W_{1\alpha}$ or $W_{1\alpha *}$ have no rapid time variation and cannot be neglected
by the above argument of time average when $m_\alpha$ satisfies the relation of $m_\alpha=m_1\pm m_{\rm D}$.
Then, the terms with $W_{1\alpha}$ or $W_{1\alpha *}$ also contribute to resonant couplings.
Such cases are outside of our present concern.
See related discussions in the final section.
}
In summary, the coupling terms remained are the middle two terms of equation (\ref{3.5}) 
with $W_{12}$ and $W_{12*}$. 

Based on the above preparations, we reduce equation (\ref{3.5}) to
\begin{eqnarray}
   &&4i\frac{E_1}{\omega_1}\frac{dA_1}{dt}
      = A_2A_{\rm D}\hat{W}_{12}{\rm exp}(-i\Delta_+ t)\delta_{m_1,m_2+m_{\rm D}}  \nonumber \\
   &&  \hspace{50pt}+ A_2A_{\rm D}^* \hat{W}_{12*}{\rm exp}(-i\Delta_- t)
        \delta_{m_1,m_2-m_{\rm D}},
\label{3.10}
\end{eqnarray}
where the symbol $\delta_{a,b}$ means that it is unity when $a=b$, but zero when $a\not= b$.
Here, from $W_{12}$ and $W_{12*}$ the time and azimuthally dependent parts are separated as
\begin{equation}
       W_{12}=\hat{W}_{12}{\rm exp}(-i\Delta_+ t)\delta_{m_1,m_2+m_{\rm D}}
\label{W12hat} 
\end{equation}
\begin{equation}
       W_{12*}=\hat{W}_{12*}{\rm exp}(-i\Delta_-  t)\delta_{m_1,m_2-m_{\rm D}}.
\label{W12*hat} 
\end{equation}
A physical meaning of equation (\ref{3.10}) is as follows.
The imaginary part of $(\omega_1/2)W_{12}$, for example, is the rate of work done on mode 1 
(when mode 2 and the deformation have unit amplitudes) through the coupling of $m_1=m_1+m_{\rm D}$
(Kato 2008a).
Hence, in a rough sense, equation (\ref{3.10}) represents the fact that the growth rate of mode 1 is
given by energy flux $F_1[=A_2A_{\rm D}(\omega_1/2)\hat{W}_{12}]$ to mode 1 as 
\begin{equation}
        \frac{dA_1}{dt}=\frac{F_1}{2E_1}.
\label{dA1overdt}
\end{equation}

The next subject is to derive an equation describing the time evolution of $A_2$ by a similar 
procedure as the above.
That is, equation (\ref{3.2}) is multiplied by $\mbox{\boldmath $\xi$}_2^*$ and integrated over
the whole volume.
Then, as the equation corresponding to equation(\ref{3.5}), we have
\begin{eqnarray}
   &&i\frac{dA_2}{dt}\frac{4E_2}{\omega_2}    \nonumber \\
   &&=A_1(A_{\rm D}W_{21}+A_{\rm D}^*W_{21*})
    +A_2(A_{\rm D}W_{22}+A_{\rm D}^*W_{22*})
    +\sum_\alpha A_\alpha(A_{\rm D}W_{2\alpha}+A_{\rm D}^*W_{2\alpha *}).
\label{3.11}
\end{eqnarray} 
Here, not all of the expressions for $W_{21}$, $W_{21*}$, $W_{22}$, $W_{22*}$, $W_{2\alpha}$, and $W_{2\alpha*}$ are given,
since they are the same as those of $W_{11}$, $W_{11*}$, $W_{12}$, $W_{12*}$, $W_{1\alpha}$, and $W_{1\alpha*}$,
respectively, except that $\mbox{\boldmath $\xi$}_1^*$ in the latters are replaced now by
$\mbox{\boldmath $\xi$}_2^*$.
As examples, we give only $W_{21}$ and $W_{21*}$ as 
\begin{equation}
       W_{21}= \frac{1}{2}\biggr(\biggr\langle\rho_0\mbox{\boldmath $\xi$}_2^*\cdot
       \mbox{\boldmath $C$}(\mbox{\boldmath $\xi$}_1, \mbox{\boldmath $\xi$}_{\rm D})\biggr\rangle
       +\biggr\langle\rho_0\mbox{\boldmath $\xi$}_2^*\cdot\mbox{\boldmath $C$}(\mbox{\boldmath $\xi$}_{\rm D}, 
                 \mbox{\boldmath $\xi$}_1)\biggr\rangle\biggr),
\label{W21}
\end{equation}
\begin{equation}
       W_{21*}= \frac{1}{2}\biggr(\biggr\langle\rho_0\mbox{\boldmath $\xi$}_2^*\cdot
       \mbox{\boldmath $C$}(\mbox{\boldmath $\xi$}_1, \mbox{\boldmath $\xi$}_{\rm D}^*)\biggr\rangle
       +\biggr\langle\rho_0\mbox{\boldmath $\xi$}_2^*\cdot\mbox{\boldmath $C$}(\mbox{\boldmath $\xi$}_{\rm D}^*, 
                 \mbox{\boldmath $\xi$}_1)\biggr\rangle\biggr).
\label{W21*}
\end{equation}

By the same argument used in reducing equation (\ref{3.5}) to equation (\ref{3.10}), 
we simplify equation (\ref{3.11}).
That is, in the present case the coupling terms that remain are those with $W_{21}$ and $W_{21*}$, and
we have finally
\begin{eqnarray}
   &&4i\frac{E_2}{\omega_2}\frac{dA_2}{dt}
      = A_1A_{\rm D}\hat{W}_{21}{\rm exp}(i\Delta_- t)\delta_{m_2,m_1+m_{\rm D}}  \nonumber \\
   &&  \hspace{50pt}+ A_1A_{\rm D}^*\hat{W}_{21*}{\rm exp}(i\Delta_+ t)
        \delta_{m_2,m_1-m_{\rm D}},
\label{3.12}
\end{eqnarray}
where $\hat{W}_{21}$ and $\hat{W}_{12*}$ are the time and azimuthal dependent parts 
of $W_{21}$ and $W_{21*}$:
\begin{equation}
       W_{21}=\hat{W}_{21}{\rm exp}(i\Delta_- t)\delta_{m_2,m_1+m_{\rm D}},
\label{W21hat} 
\end{equation}
\begin{equation}
       W_{21*}=\hat{W}_{21*}{\rm exp}(i\Delta_+ t)\delta_{m_2,m_1-m_{\rm D}}.
\label{W21*hat} 
\end{equation}
The main results obtained in this section are equations (\ref{3.10}) and (\ref{3.12}).

Here, it is of importance to note that we have the following identical relations:
\begin{equation}
      W_{21}=(W_{12*})^*,
\label{3.14}
\end{equation}
\begin{equation}
      W_{21*}=(W_{12})^*,
\label{3.15}
\end{equation}
where the superscript * means the complex conjugate.
These relation come from the fact that for arbitrary functions, $\mbox{\boldmath $\eta$}_1$,
$\mbox{\boldmath $\eta$}_2$, and $\mbox{\boldmath $\eta$}_3$, their order in 
$\langle\rho_0\mbox{\boldmath $\eta$}_1\cdot\mbox{\boldmath $C$}(\mbox{\boldmath $\eta$}_2, \mbox{\boldmath $\eta$}_3)\rangle$
can be arbitrary changed [see equation (\ref{commutative})] (Kato 2008a).

\section{Growth Rate of Resonant Oscillations}

By solving the set of equations (\ref{3.10}) and (\ref{3.12}), we examine how amplitudes of $A_1$ and $A_2$ evolve with time.
We consider two cases of $m_2=m_1+m_{\rm D}$ and $m_2=m_1-m_{\rm D}$, separately.

\subsection{Case of $m_2=m_1+m_{\rm D}$}

In this case the set of equations of $A_1$ and $A_2$ are, from equations (\ref{3.10}) and (\ref{3.12}),
\begin{equation}
     4i\frac{E_1}{\omega_1}\frac{dA_1}{dt}= A_2A_{\rm D}^*\hat{W}_{12*}{\rm exp}(-i\Delta_-t),
\label{4.1}
\end{equation}
\begin{equation}
     4i\frac{E_2}{\omega_2}\frac{dA_2}{dt}= A_1A_{\rm D}\hat{W}_{21}{\rm exp}(i\Delta_-t).
\label{4.2}
\end{equation}
By introducing a new variable ${\tilde A}_1$ defined by
\begin{equation}
      {\tilde A}_1=A_1{\rm exp}(i\Delta_- t),
\label{4.4}
\end{equation}
we can reduce the above set of equations to
\begin{equation}
     4i\frac{E_1}{\omega_1}\frac{d{\tilde A}_1}{dt}+4\frac{E_1}{\omega_1}\Delta_- {\tilde A}_1
         =A_2A_{\rm D}^*W_{12*},
\label{4.4'}
\end{equation}
\begin{equation}
    4i\frac{E_2}{\omega_2}\frac{dA_2}{dt}={\tilde A}_1A_{\rm D}\hat{W}_{21}.
\label{4.5}
\end{equation}
Hence, by taking ${\tilde A}_1$ and $A_2$ to be proportional to ${\rm exp}(i\sigma t)$,
we obtain an equation describing $\sigma$ as
\begin{equation}
     \sigma^2-\Delta_-\sigma-\frac{\omega_1\omega_2}{16E_1E_2}\vert A_{\rm D}\vert^2 \vert W_{12*}^*\vert^2=0,
\label{4.6}
\end{equation}
where equation(\ref{3.14}) is used.

In the limit of an exact resonance of $\Delta_-=0$, the instability condition ($\sigma^2<0$) 
is found to be $(\omega_1/E_1)(\omega_2/E_2)<0$.
The meaning of this condition is discussed later.
If the frequencies of two oscillations deviate from the resonant condition of
$\omega_1=\omega_2-\omega_{\rm D}$, the growth rate decreases.
This can be shown from equation (\ref{4.6}).
That is, the condition of growth is
\begin{equation}
     \Delta_-^2+\frac{\omega_1\omega_2}{4E_1E_2}\vert A_{\rm D}\vert^2 \vert W_{12*}\vert^2<0,
\label{4.7}
\end{equation}
and the growth rate tends to zero, as $\Delta_-^2$ increases from zero.
If $\Delta_-^2$ increases beyond a certain limit the left-hand side of inequality (\ref{4.7}) becomes positive, and
$\sigma$ is no longer complex.
That is, the amplitude of oscillations are modulated with
time, but there is no secular increase of them.

\subsection{Case of $m_2=m_1-m_{\rm D}$}

In the present case, from equations (\ref{3.10}) and (\ref{3.12}), we have
\begin{equation}
   4i\frac{E_1}{\omega_1}\frac{dA_1}{dt}=A_2A_{\rm D}\hat{W}_{12}{\rm exp}(-i\Delta_+ t),
\label{4.8}
\end{equation}
\begin{equation}
   4i\frac{E_2}{\omega_2}\frac{dA_2}{dt}=A_1A_{\rm D}^*\hat{W}_{21*}{\rm exp}(i\Delta_+ t).
\label{4.9}
\end{equation}
A new variable ${\tilde A}_1$ is introduced here by 
\begin{equation}
      {\tilde A}_1=A_1{\rm exp}(i\Delta_+ t).
\label{4.10}
\end{equation}
Then, the set of equations (\ref{4.8}) and (\ref{4.9}) are reduced to a set of equations of 
${\tilde A}_1$ and $A_2$ as
\begin{equation}
      4i\frac{E_1}{\omega_1}\frac{d{\tilde A}_1}{dt}+\frac{4E_1}{\omega_1}\Delta_+{\tilde A}_1 
      = A_2A_{\rm D}\hat{W}_{12},
\label{4.11}
\end{equation}
\begin{equation}
    4i\frac{E_2}{\omega_2}\frac{dA_2}{dt}={\tilde A}_1A_{\rm D}^*\hat{W}_{21*}.
\label{4.12}
\end{equation}
Hence, by taking ${\tilde A}_1$ and $A_2$ to be proportional to ${\rm exp}(i\sigma t)$,
we have
\begin{equation}
    \sigma^2-\Delta_+\sigma-\frac{\omega_1\omega_2}{16E_1E_2}\vert A_{\rm D}\vert^2 
        \vert\hat{W}_{12}\vert^2=0,
\label{4.13}
\end{equation}
where we have used equation (\ref{3.15}).

Two oscillations certianly grow again at the limit of exact resonance of $\Delta_+=0$
(i.e., $\omega_1=\omega_2+\omega_{\rm D}$), if $(\omega_1/E_1)(\omega_2/E_2)<0$.
Even if the resonance is not exact, they grow if $\Delta_+^2$ is small enough so that
\begin{equation}
     \Delta_+^2+\frac{\omega_1\omega_2}{4E_1E_2}\vert A_{\rm D}\vert^2
       \vert\hat{W}_{12}\vert^2<0
\label{4.14}
\end{equation}
is satisfied.

\subsection{A Relation between $A_1$ and $A_2$}

Finally, it is useful to derive an instructive relation between $A_1$ and $A_2$.
Let us first consider the case of $m_2=m_1+m_{\rm D}$.
Let us multiply $A_1^*$ to equation (\ref{4.1}) and also $A_1$ to the complex conjugate 
of equation (\ref{4.1}).
Then, summing these two equations we have an equation describing time evolution of
$\vert A_1\vert^2$.
Similarly, from equation (\ref{4.2}), we can derive an equation describing time evolution of
$\vert A_2\vert^2$.
Summing these two equations, we have finally
\begin{equation}
    \frac{d}{dt}\biggr[\frac{E_1}{\omega_1}\vert A_1\vert^2+
    \frac{E_2}{\omega_2}\vert A_2\vert^2\biggr]=0,
\label{cons}
\end{equation}
where we have used $\hat{W}_{21}^*=\hat{W}_{12*}$.
The same equation can be derived from euqations (\ref{4.8}) and
(\ref{4.9}) in the case of $m_2=m_1-m_{\rm D}$.
To derive the equaition, $\hat{W}_{21*}=\hat{W}_{12}^*$ has been used.

Equation (\ref{cons}) obviously shows that $(\omega_1/E_1)(\omega_2/E_2) <0$ is necessary for growth 
of oscillations.
In the case of $(\omega_1/E_1)(\omega_2/E_2) >0$, on the other hand, amplitudes of $A_1$ and $A_2$ are limited, 
although the relative amplitude of both oscillations may change with time by
interaction through disk deformation.

\subsection{Summary of Resonant Instability Condition}

The results in the previous subsections show that when resonant conditions of $\omega_1=\omega_2\pm \omega_{\rm D}$ and 
$m_1=m_2\pm m_{\rm D}$ are satisfied among two oscillations characterized by ($\omega_1$, $m_1$) and ($\omega_2$, $m_2$)
and disk deformation characterized by ($\omega_{\rm D}$, $m_{\rm D}$),
the two oscillations are resonantly excited if $(\omega_1/E_1)(\omega_2/E_2)<0$ is realized.
A deviation from the condition of $\omega_1=\omega_2\pm \omega_{\rm D}$ decreases the growth rate, but oscillations grow 
as long as the deviation is smaller than a critical value.

In the case where both of $\omega_1$ and $\omega_2$ are positive (i.e., both oscillations are prograde),
the above instability condition is $E_1E_2<0$.
This is the result suggested by Kato (2004, 2008a, b) by a different approach.

In the case where $\omega_{\rm D}$ is larger than $\omega_2(>0)$, a resonant condition, $\omega_1=\omega_2-\omega_{\rm D}$,
is satisfied for $\omega_1<0$ (i.e., retrograde wave).
In this case the wave energy $E_1$ is positive [see equation (\ref{energy})] and the instability condition is reduced to
$E_2/\omega_2>0$, i.e., $E_2>0$.
That is, when $\omega_1\omega_2<0$, the condition of resonant instability is $E_1E_2>0$.
It is noted that in the case where both of $\omega_1$ and $\omega_2$ are negative, both $E_1$ and $E_2$ are positive, so that  
the condition of resonant instability, $(\omega_1/E_1)(\omega_2/E_2)<0$, cannot be satisfied.

Among three case of i) $\omega_1>0$ and $\omega_2>0$, ii) $\omega_1<0$ and $\omega_2>0$, and iii) $\omega_1<0$ and $\omega_2<0$,
the interesting case in the practical sense is the first one, which is discussed in the next section.

\section{Discussion}

First, let us describe, in terms of the present formulation, the g- and p-modes resonant instability that was numerically
studied by Ferreira and Ogilvie (2008) and Oktariani et al. (2010).
They considered the resonant interaction, through a standing warp ($\omega_{\rm D}=0$), 
between i) the axisymmetric g-mode oscillation whose 
$\xi_r$ has one node in the vertical direction and ii) the one-armed p-mode oscillation 
whose $\xi_r$ has no node in the vertical direction.
That is, the set of ($\omega$, $m$, $n$) is ($\sim\kappa_{\rm max}$, 0, 1) for the g-mode oscillation,
and ($\sim\kappa_{\rm max}$, 1, 0) for the p-mode one, where
$\kappa_{\rm max}$ is the maximum of the (radial) epicyclic frequency.
The warp is taken to be (0, 1, 1).
In this case, the resonant conditions, $\omega_1\sim\omega_2>0$ (i.e., $\omega_{\rm D}=0$) and $m_2=m_1+m_{\rm D}$,
are satisfied.
Hence, if $E_1E_2<0$, these modes are excited simultaneously.
This condition of $E_1E_2<0$ is really satisfied, since the axisymmetric g-mode oscillation has a positive energy, 
while the p-mode oscillation 
trapped between the inner edge of the disk and the barrier resulting from the boundary between the propagation and 
evanescent regions has a negative energy.
The cause of excitation is a positive energy flow from a negative energy oscillation (p-mode) to a positive
energy oscillation (g-mode).
By this energy flow both oscillations grow.
The disk deformation is a catalyzer of this energy flow.

In the above argument the disk deformation is assumed to be a warp.
Instead of a warp, we can consider a c-mode oscillation as one of other possible disk deformations.
In this case the set of ($\omega$, $m$, $n$) of the disk deformation is ($\omega_{\rm D}$, 1, 1), where $\omega_{\rm D}$
is the frequency of c-mode oscillation and $\omega_{\rm D}\ll \kappa_{\rm max}$, unless the spin
of the central source is high (Silbergleit et al. 2001).
The resonant condition in this case is $\omega_1=\omega_2\pm \omega_{\rm D}$, not $\omega_1=\omega_2$.

There are some limitations of a direct comparison of the present analytical results to the numerical ones by
Ferreira and Ogilvie (2008).
In the present analyses only two normal modes of oscillations are assumed to contribute to the resonance to understand the essence 
of the resonant instability.
In the realistic case considered numerically by Ferreira and Ogilvie (2008), however, more than two normal modes of oscillations
may contribute to the resonance.
In their case one of resonant oscillations is an axisymmetric g-mode.
As mentioned in footnote 4, eigen-frequencies of the axisymmetric ($m=0$) g-mode oscillations with different $\ell$ 
(with $n=1$) are all close to $\kappa_{\rm max}$.
Hence, the g-mode oscillations that satisfy the resonant conditions, $\omega_1=\omega_2+\omega_{\rm D}$ and $m_1=m_2+m_{\rm D}$, 
may not be only one, and overtones of g-mode oscillations
with nodes in the radial direction may also contribute partially to the resonance.
If this is the case, some coupling terms with other $A$ than $A_1$ and $A_2$ appear in equations describing time evolution of
$A_1$ and $A_2$. 
In the present analytical formulation such situations are not considered.
To extend our analyses to such cases, we must derive equations describing the time evolution of other $A$'s than $A_1$ and $A_2$,
and these equations should be solved simultaneously with the equations describing time evolution of $A_1$ and $A_2$.
We think that the essence of the instability mechanism is already presented in the case where couplings occur only
between two oscillations.
However, since such extention of our formulation is formally simple and there may be some subsidiary modifications
of instability criterion, such extension should be done in the near future.  

In the present formulation, some important theoretical problems remain to be clarified.
One of them is whether the set of normal modes of oscillations form a complete set.
If not, it is uncertain whether the oscillations realized on the disk can be expressed in the form of
equation (\ref{3.1}).

In the case where $\omega_{\rm D}$ is larger than $\omega_2$, the resonant frequency $\omega_1$ which satisfies the 
condition, $\omega_1=\omega_2-\omega_{\rm D}$, is negative.
In this case of $\omega_1<0$, $E_1$ is positive [see equation (\ref{energy})].
Hence, the instability condition in this case is $E_1E_2>0$, as mentioned in the last section.
All the g-mode and p-mode oscillations with negative frequency are, however,  not trapped in the inner region of disks.
They propagate away far outside unless the disks are truncated,\footnote{
This can be shown by examining local dispersion relation of these oscillations.
The vertical p-mode oscillations that are trapped in finite region also have $\omega_1>0$.
}
That is, their frequencies are continuous and will not be directly related to the QPO phenomena in disks. 

It is important to note here that the innermost region of relativistic disks is a place where the present 
excitation mechanism works most efficiently.
For the mechanism to work, two normal modes of oscillations with opposite signs of wave energy must 
coexist in a common region of disks.
In general, for $m\ne 0$, positive energy oscillations propagate in the region outside the radius of corotation resonance, 
while negative energy oscillations inside the resonance.
Hence, there is a tendency that the propagation region of oscillations with opposite signs of wave energy
are spatially separated, unless different types of oscillation modes are considered.
If the propagation region of oscillations with opposite signs of wave energy are separated, 
the coupling efficiency between the two
oscillations are weak, i.e., $W$'s in equations (\ref{3.10}) and (\ref{3.12}) are small, and
practically there is no growth of oscillations.
For oscillations with opposite signs of wave energy to coexist in a common region, effects of general 
relativity is important.
It is noted that in such cases there is a tendency that the Lindblad resonance of an oscillation (e.g., p-mode)
occurs in the propagation region of the other oscillation (e.g., g-mode).
This helps to increase the growth rate since the coupling terms become large, as mentioned before.

Finally, as an application of the present resonant excitation process of oscillations, we
briefly note Kato's model of high-frequency twin QPOs (e.g., Kato and Fukue 2006).
In this model the lower-frequency QPO of the twin is related to the set of g-mode and p-mode oscillations
considered numerically by Ferreira and Ogilvie (2008) and Oktariani et al. (2010).
The higher-frequency QPO of the twin is considered to be the set of  
one-armed g-mode and two-armed p-mode oscillations [see for details figures
1 to 3 of Kato (2008a)].
Here, the g-mode oscillation has a positive energy and the p-mode one has a negative energy,
and thus this set of oscillations can also grow by the resonant coupling.

In his model, correlated time variation of twin QPOs observed in neutron-star low-mass X-ray binaries are
described by assuming time variation of $\omega_{\rm D}$.
One of problems of this model, however, is that masses of neutron stars required to take into account observations
are rather high, i.e., for example, 2.4 $M_\odot$ for Sco X-1 and 4U 1636-53.
See Lin et al. (2010) for detailed comparison of various QPO models with observations.
In black-hole low-mass X-ray binaries, the twin QPOs have no frequency change with frequency ratio 
of 3 : 2.
In Kato's model this can be described by assuming $\omega_{\rm D}=0$.
The spin parameters $a_*$'s estimated by this model for black hole sources with measured masses are around
$a_*\sim 0.4$ (e.g., Kato and Fukue 2006).
However, the spin parameters estimated by comparing model continuum X-ray spectra with observations
are generally higher than the above, say around $a_*\sim 0.8$ or more (e.g., Narayan et at. 2008).

\bigskip
\leftskip=20pt
\parindent=-20pt
\par
{\bf References}
\par
Ferreira, B.T. and Ogilvie, G.I. 2008, MNRAS, 386, 2297 \par
Kato, S. 2001, PASJ, 53, 1\par 
Kato, S. 2004, PASJ, 56, 905\par
Kato, S. 2008a, PASJ, 60, 111 \par
Kato, S. 2008b, PASJ, 60, 1387 \par
Kato, S. 2009, PASJ, 61, 1237\par
Kato, S. and Fukue, J. 2006, PASJ, 58, 909 \par  
Kato, S., Fukue, J., \& Mineshige, S. 2008, Black-Hole Accretion Disks 
  -- Toward a New paradigm -- (Kyoto: Kyoto University Press)\par
Lin, Y.F., Boutelier, M., Barret, D., \& Zhang, S.N. 2010, Apj in press, astro-ph. arXiv: 1010.6198\par
Lynden-Bell, D. and Ostriker, J.P. 1967, MNRAS, 136, 293  \par
Narayan, R., McClintock, J.E., \& Shafee, R. 2008, in Astrophysics of Compact Objects, eds.
    Y.F. Yuan, X.D. Li, \& D.Lai, AIP Conf. Proc. 968, 265\par
Okazaki, A.-T., Kato, S., and Fukue, J. 1987, PASJ, 39, 457 \par
Oktariani, F., Okazaki, A.-T., \& Kato, S. 2010, PASJ, 62, 709 \par
Paczy\'{n}ski, B.,\& Wiita, P>J. 1980, A\&A 88, 23\par
Silbergleit, A.S., Wagoner, R., Ortega-Rodriguez, M. 2001, APJ, 548, 335\par   
\leftskip=0pt
\parindent=20pt

\end{document}